# Cathodoluminescence and Selective Emission of Er$^{3+}$ in Oxides

V. M. Marchenko, M. G. Voitik, and V. A. Yuryev

*General Physics Institute, Russian Academy of Sciences, ul. Vavilova 38, Moscow, 119991 Russia*
e-mail: vmarch@kapella.gpi.ru

**Abstract**—The dependence of the Er$^{3+}$ cathodoluminescence and selective emission on the power of the YAG:Er$^{3+}$ and Er$_2$O$_3$ excitation by an electron beam is spectroscopically studied for applications in high-intensity radiation sources of the visible and near-IR spectral ranges.
**DOI:** 10.1134/S1054660X0806

(i) Rare-earth oxides (in particular, Er$_2$O$_3$) are used as thermally stable selective emission (SE) sources in thermophotovoltaic (TPV) generators [1–3]. The thermal excitation of the vibronic spectra of the Er$_2$O$_3$ SE results from multiphoton transitions from the ground to excited electronic states of the screened $4f^{11}$ electronic shells of Er$^{3+}$ and the cross relaxation [4–6]. A relatively high SE integral intensity in the visible and near-IR spectral ranges has been realized upon the laser thermal excitation of Er$_2$O$_3$ at a wavelength of 10.6 μm [6]. The resonant optical pumping of the active media activated with rare-earth ions causes an increase in the laser efficiency for the visible and near-IR spectral ranges [7–12]. This accounts for the interest in the development of the thermophoto-lasers (TPLs) [6, 13, 14] whose laser elements are pumped by the thermally non-equilibrium SE [15]. The electron-beam excited SE of the rare-earth oxides can be another variant of the optical pumping.

(ii) In this work, we experimentally study the dependence of the emission spectra and intensity of YAG:Er$^{3+}$ and Er$_2$O$_3$ oxides on the electron-beam-excitation power. The oxide samples placed on the anode are irradiated over several minutes using a beam of the runaway electrons in the electric discharge at the voltage $U$ = 2–6 kV between the tungsten cathode and the tantalum anode in a quartz tube (with a diameter of 1 cm) filled with helium or nitrogen at the pressure $p$ = 0.05– 0.2 Torr. In this pressure range, we can neglect the electron-energy loss $\Delta E \approx xNL(p)$ ($x \approx$ 3 cm is the electron-path length, $N$ is the molecular concentration, and $L(p) \approx 2 \times 10^{-16}$ eV cm$^2$ is the electron scattering function in helium [16]) and can assume with an accuracy sufficient for estimations that the electron energy is $eU \sim$ 2–6 kV ($e$ is elementary charge) in the vicinity of the target.

The parent samples are the YAG:Er$^{3+}$ laser single crystal and the Er$_2$O$_3$ crystalline powder with a dispersity of 5–30 μm. The YAG:Er$^{3+}$ and Er$_2$O$_3$ polycrystalline samples are synthesized using the melting of powders irradiated with a CO$_2$ laser at a radiation power of about 100 W and a wavelength of 10.6 μm on a substrate made of single-crystal silicon. The dispersed oxides are produced using the crushing and grinding of the single-crystalline and polycrystalline powders in a mortar.

The electron-beam excited emission spectra of the oxides $I(\lambda) \sim n(\lambda)/\Delta\lambda$ ($n(\lambda)$ is the number of photons in the spectral interval $\Delta\lambda$) are measured using an FSD-6 diffraction spectrometer with a fiber-optic input directed at a certain angle relative to the electron-irradiated oxide target. The measurements are performed in the sensitivity spectral range (300–900 nm) of a semiconductor detector array with the resolution $\Delta\lambda$ = 10 nm. The normal scan duration is 40 ms, and the number of scans is about 10. Using computer processing, we perform the correction of the measured emission spectra with regard to the spectral sensitivity of the array, the smoothing of the noisy spectra, and the intensity normalization needed for the curve plotting. Integral-spectral intensity $I_\Sigma$, which is proportional to the total number of photons in the sensitivity spectral range, is calculated for the spectral series measured at the given experimental configuration [6].

(iii) The typical dependences of the normalized emission spectra $I_n(\lambda) = I(\lambda)/I_{max}(\lambda)$ of the YAG:Er$^{3+}$ single crystal, crystalline powder, and polycrystal and the Er$_2$O$_3$ powder on the electron-beam excitation power $P = iU$ are presented in Fig. 1. Table 1 shows the discharge-tube voltages $U$ and currents $i$. On the plots, the spectrum numbers correspond to the column numbers in the first row and the terms of the electronic states involved in the radiative transitions to the ground state are denoted in accordance with the interpretation from [4]. For the YAG:Er$^{3+}$ single crystal, the emission spectra measured in the above range of the electron-beam excitation power (Fig. 1a) coincide with the cathodoluminescence (CL) spectra [17–20].

The curve of the CL integral intensity of the YAG:Er$^{3+}$ single crystal $I_\Sigma(P)$ is linear at $P \leq 20$ W (Fig. 2a). The relative intensity corresponding to the measured short-wavelength wing of the $^4I_{11/2}$ term of the Er$^{3+}$ ion decreases with an increasing $P$ in the linear range of the dependence $I_\Sigma(P)$ (Fig. 1a and Table 1a). This is due to the temperature dependence of the spectra and the upconversion processes involving the excitation of higher energy levels [4, 6]. The saturation of the dependence $I_\Sigma(P)$ is realized at $P > 20$ W.

The power dependences (Tables 1b and 1c) of the spectra $I_n(\lambda)$ of the YAG:Er$^{3+}$ polycrystals (Fig. 1b) and the powders with different dispersities produced using different methods (Fig. 1c) appear similar. However, these dependences differ from the dependence obtained for single crystals in spite of the similar curves of the integral intensity $I_\Sigma(P)$ (Figs. 2b and 2c). At relatively low $P$, the spectra of the polycrystals and powders are similar to the spectra of the single crystal (spectra *1* in Figs. 1b and 1c). When $P$ increases, the relative intensity of the short-wavelength spectral components ($^4F-^4G)_{9/2}-^4F_{9/2}$, first, increases as in the case of the YAG:Er$^{3+}$ single crystal (spectra *2* in Figs. 1b and 1c) and, then, decreases, so that the relative intensity of the long-wavelength spectral components $^4I_{9/2}$ rapidly increases (spectra *3* and *4* in Fig. 1b and spectrum *3* in Fig. 1c). At a relatively high power, the spectral selectivity decreases and the relative intensity of the short-wavelength spectral components become negligible (spectrum *5* in Fig. 1b and spectrum *4* in Fig. 1c). The same regularity lying in an increase in the relative intensity of the long-wavelength components and a decrease in the spectral selectivity with an increase in the electron-beam excitation power is characteristic of the transformation of the Er$_2$O$_3$-batch emission spectra (Fig. 1d and Table 1d). Such an increase in the relative intensity of the long-wavelength components for vibronic spectra of the Er$^{3+}$ SE and $I_\Sigma(P)$ of the Er$_2$O$_3$ polycrystals and powders with an increase in the intensity of the laser thermal excitation at a wavelength of 10.6 µm can be found in [6]. Figure 3a and Table 2 illustrate the difference in the spectral shapes for the mission of the YAG:Er$^{3+}$ single crystal and the Er$_2$O$_3$ using the electron-thermal and laser-thermal excitations are similar (Fig. 3b).

(iv) Thus, an increase in the relative intensity of the long-wavelength components in the spectra of YAG:Er$^{3+}$ and Er$_2$O$_3$ polycrystals and powders with an increase in the electron-irradiation power is caused by both CL and the thermal SE, which involves an additional electron-thermal mechanism for the excitation of the corresponding electronic states of Er$^{3+}$. The difference between the excitation-intensity dependences and selectivities of the CL of the YAG:Er$^{3+}$ single crystal and the SE of the YAG:Er$^{3+}$ polycrystals and powders at a relatively high excitation power must be due to the difference between the temperatures of the samples related to significantly different thermal conductivities and due to the SE thermalization. A relatively high SE efficiency proved by the linear increase of $I_\Sigma(P)$ upon a decrease in the CL relative intensity serves as the basis for the high-intensity electron-beam radiation sources of the visible and IR spectral ranges needed for various practical applications.

REFERENCES

1. D. L. Chubb, A.-M. T. Pal, M. O. Patton, and P. P. Jenkins, J. Eur. Ceram. Soc. **19**, 2551 (1999).
2. B. Bitnar, W. Durisch, J.-C. Mayor, et al., Sol. Energy Mater. Sol. Cells **73**, 221 (2002).
3. A. Licciulli, D. Diso, G. Torsello, et al., Semicond. Sci. Technol. **18**, S174 (2003).
4. X. S. Bagdasarov, V. I. Zhekov, V. A. Lobachev, et al., "Cross-Relaxation YAG-Er3-Laser," Tr. Inst. Obshch. Fiz., Ross. Akad. Nauk **19**, 5 (1989).
5. G. Torsello, M. Lomascollo, A. Licciulli, et al., Nature Mater. **3**, 632 (2004).
6. V. M. Marchenko, Laser Phys. **17**, 1146 (2007).
7. T. J. Whitley, Electron. Lett. **25**, 1537 (1988).
8. M. Tikerpae, S. D. Jackson, T. A. King, J. Mod. Opt. **45**, 1269 (1998).
9. A. M. Tkachuk, I. K. Razumova, A. A. Mirzaeva, et al., Opt. Spektrosk. **92**, 73 (2002) [Opt. Spectrosc. **92**, 67 (2002)].
10. E. Dianov, Usp. Fiz. Nauk **174**, 1139 (2004).
11. V. P. Danilov, B. I. Denker, V. I. Zhekov, et al., Pis'ma Zh. Tekh. Fiz. **32**, 40 (2006).

12. Shilong Zhao, Shunguang Li, Lili Hu, and Zaixuan Zhang, Chin. Opt. Lett. **4**, 228 (2006).
13. D. L. Ghubb and M. O. Patton, US Patent 6 198 760 (1996).
14. V. M. Marchenko, Quantum Electron. **36**, 727 (2006).
15. L. D. Landau and E. M. Lifshits, *Statistical Physics* (Nauka, Moscow, 1964; Pergamon, Oxford, 1980).
16. V. I. Kolobov and L. D. Tsendin, Phys. Rev. A **46**, 7837 (1992).
17. R. Plugaru, J. Piqueras, E. Nogales, et al., J. Optoelectron. Adv. Mater. **4**, 883 (2002).
18. E. Nogales, B. Mendez, J. Piqueras, et al., J. Phys. D: Appl. Phys. **35**, 295 (2002).
19. W. M. Jadwisienczak, H. J. Lozykowski, A. Xu, and B. Patel, J. Electron. Mater. **31**, 776 (2002).
20. A. V. Rasuleva and V. I. Solomonov, Laser Phys. **16**, 130 (2006).

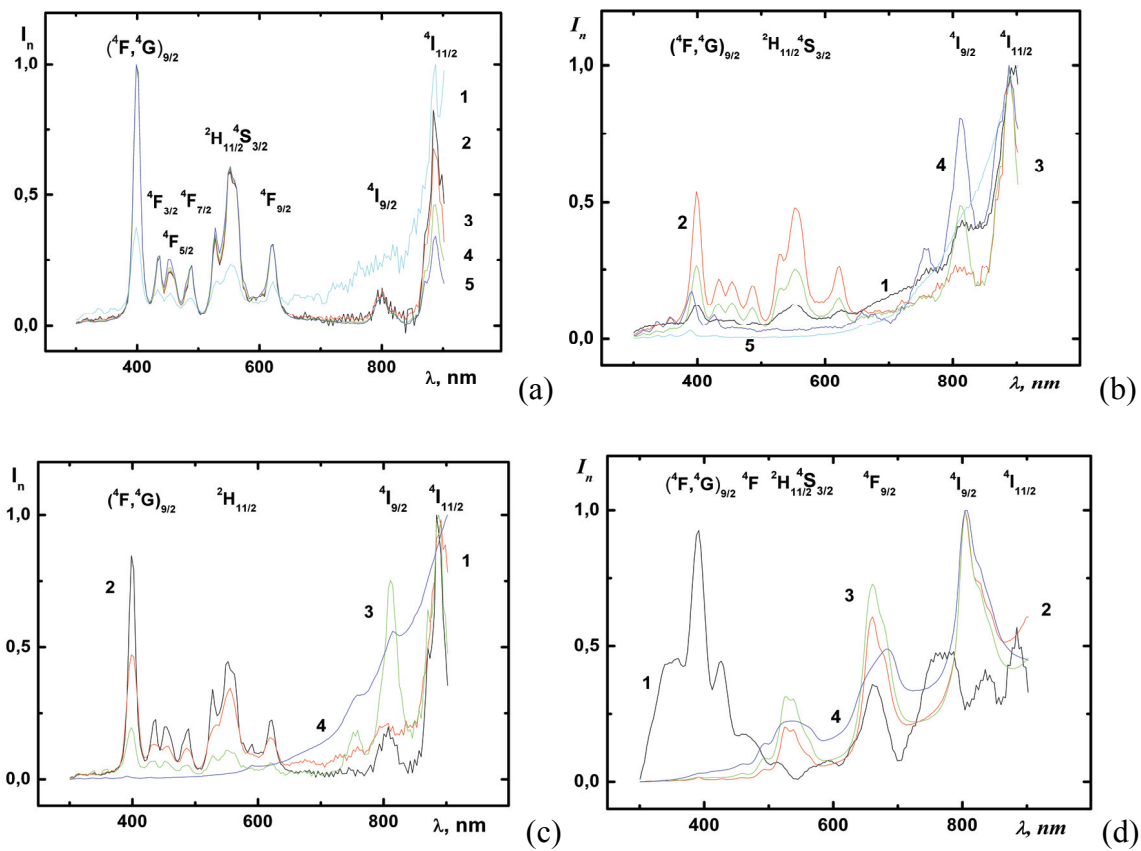

**Fig. 1.** The electron-beam excited emission spectra of (a) YAG:Er$^{3+}$ single crystal, (b) YAG:Er$^{3+}$ polycrystal, (c) YAG:Er$^{3+}$ crystalline powder, and (d) Er$_2$O$_3$ powder. For each plot, the discharge-tube voltages and currents corresponding to the spectrum numbers are presented in Table 1.

**Table 1.**

| a | | | | | | b | | | | | |
|---|---|---|---|---|---|---|---|---|---|---|---|
| No. | 1 | 2 | 3 | 4 | 5 | No. | 1 | 2 | 3 | 4 | 5 |
| $U$, kV | 3 | 3,6 | 4 | 4,6 | 5,5 | $U$, kV | 6 | 4 | 6 | 6 | 6 |
| $i$, mA | 0,4 | 0,8 | 1,4 | 2,5 | 4,1 | $i$, mA | 0,3 | 4,2 | 3 | 12 | >12 |

| c | | | | | d | | | | |
|---|---|---|---|---|---|---|---|---|---|
| No. | 1 | 2 | 3 | 4 | No. | 1 | 2 | 3 | 4 |
| $U$, kV | 2,4 | 3 | 4 | >4 | $U$, kV | 4 | 5 | 6 | 6 |
| $i$, mA | 0,35 | 1 | 3,6 | - | $i$, mA | 3,6 | 9 | 11 | >11 |

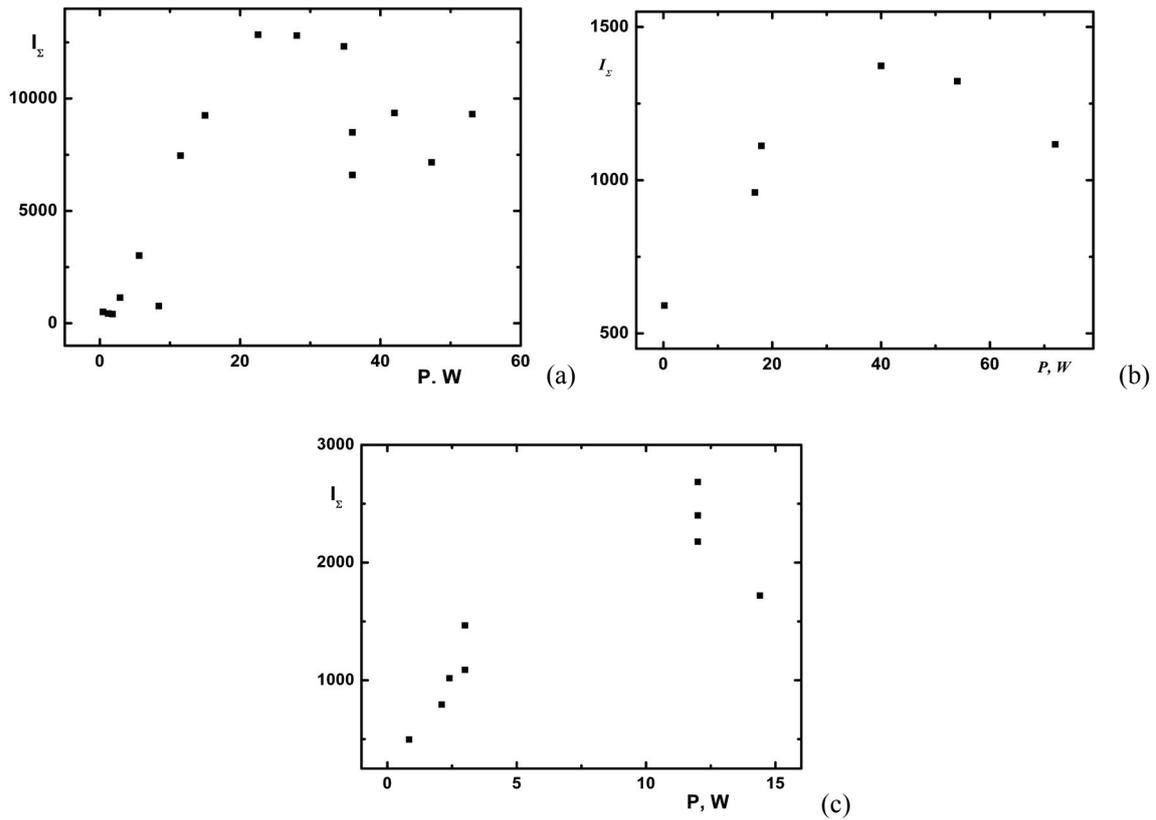

**Fig. 2.** Plots of the integral-emission intensity $I_\Sigma$ of the YAG:Er$^{3+}$ (a) single crystal, (b) polycrystal, and (c) powder vs. electron-beam power $P$.

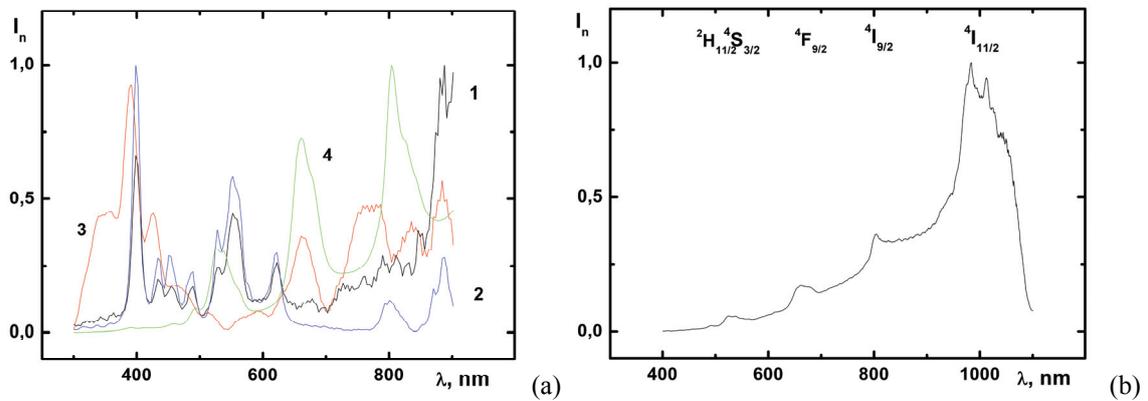

**Fig. 3.** (a) The electron-beam excited emission spectra of (*1*) and (*2*) the YAG:Er$^{3+}$ single crystal and (*3*) and (*4*) the Er$_2$O$_3$ powder. (b) The emission spectrum of the Er$_2$O$_3$ powder upon the laser-thermal excitation. The discharge-tube voltages and currents corresponding to the spectrum numbers are presented in Table 2.

**Table 2.**

| No. | 1 | 2 | 3 | 4 |
|---|---|---|---|---|
| $U$, kV | 3 | 5,9 | 4 | 5 |
| $i$, mA | 0,4 | 9 | 3,6 | 9 |